\documentclass[a4paper]{article}

\usepackage[utf8]{inputenc}
\usepackage[round]{natbib}
\usepackage[english]{babel}
\usepackage{amsfonts}
\usepackage{amsmath}
\usepackage{amsthm}
\usepackage{amssymb}
\usepackage{dsfont}
\usepackage{url}
\usepackage{hyperref}
\usepackage{tikz-cd}
\usepackage{mathtools}
\usepackage{enumitem}
\usepackage{color}
\usepackage{mathpartir}
\usepackage{stmaryrd}
\usepackage{listings}
\usepackage{listings-rust}

\mathtoolsset{showonlyrefs}

\theoremstyle{remark}

\theoremstyle{definition}

\begin{document}

\title{An Evaluation Algorithm for Datalog with Equality}

\author{Martin E. Bidlingmaier}

\date{}

\maketitle

\begin{abstract}
  We describe an evaluation algorithm for \emph{relational Horn logic (RHL)}.
  RHL extends Datalog with quantification over sorts, existential quantification in conclusions and, crucially, the ability to infer \emph{equalities}.
  These capabilities allow RHL evaluation to subsume and expand applications of Datalog and congruence closure algorithms.

  We explain how aspects of a fast congruence closure algorithm can be incorporated into Datalog evaluation to obtain an efficient RHL evaluation algorithm.
  We then sketch how Steensgaard's points-to analysis and type inference can be implemented using RHL evaluation.
  RHL and the evaluation algorithm described here are the foundation of the \emph{Eqlog} Datalog engine.
\end{abstract}

\section{Introduction}

Recent work has identified \emph{relational Horn logic} (RHL) and \emph{partial Horn logic} (PHL) as a semantically well-behaved extensions of Datalog \citep{phl-theory}.
This paper describes how the Datalog evaluation algorithm can be generalized to RHL evaluation.

An RHL theory contains declarations of the following data:
\begin{itemize}
  \item
    A set $S$ of \emph{sorts}.
  \item
    A set $R$ of \emph{relations}.
  \item
    \emph{Arities} $r : s_1 \times \dots \times s_n$ for all relations $r \in R$.
  \item
    A set of \emph{sequents} (or \emph{rules}, or \emph{axioms}), each of the form $\mathcal{F} \implies \mathcal{G}$, where $\mathcal{F}$ and $\mathcal{G}$ are conjunctions $\phi_1 \land \dots \land \phi_n$ of \emph{atoms} $\phi_i$.
\end{itemize}
Instead of RHL sequents as above, Datalog engines typically accept rules of the following form:
\begin{equation}
  \phi \; \text{:-} \; \phi_1, \dots, \phi_n.
\end{equation}
The \emph{head} $\phi$ corresponds to the conclusion of an RHL sequent and consists of a single atom.
The \emph{body} $\phi_1, \dots, \phi_n$ corresponds to the premise of an RHL sequent, can contain multiple atoms and is interpreted as conjunction.
The structure of RHL sequents is thus more general at first sight because the conclusion of an RHL sequents is allowed to be a conjunction of atoms.
However, a single sequent with $n$ conclusion atoms is equivalent to $n$ sequents, each with a single conclusion atom.
Thus, no generality is gained solely from allowing general conjunctions as conclusions.

Where RHL generalizes Datalog is in what kind of atoms it allows, and how variables are handled.
In Datalog, each atom is of the form $r(v_1, \dots, v_n)$ where $r$ is a relation symbol and the $v_i$ are variables whose sort match the arity of $r$.
In addition to such \emph{relation atoms}, RHL recognizes also the following types of atoms:
\begin{enumerate}
  \item
    An \emph{equality atom}
    \begin{equation}
      u \equiv v
    \end{equation}
    where $u$ and $v$ have the same sort.
    We reserve the symbol ${\equiv}$ for RHL syntax, whereas ${=}$ is used for meta-theoretical equality.
  \item
    A \emph{sort quantification}
    \begin{equation}
      v \downarrow
    \end{equation}
    where $v$ is a variable with known sort $s$.
    If the sort of $v$ is not determined by other atoms in the sequent, we also use the syntax $v : s$ as synonymous for $v \downarrow$ and the metatheoretical assertion that $v$ has sort $s$.
\end{enumerate}
If an equality atom $u \equiv v$ occurs in the premise of a sequent, then matches of the premise are only valid if $u$ and $v$ are interpreted as the same constant.
Thus, equality atoms in a premise can be eliminated by removing all occurrences of one of the two variables in the sequent by the other variable.

The semantics of an equality atom $u \equiv v$ in the conclusion of a sequent are non-trivial, however:
Whenever the premise of such a sequent matches such that $u$ is interpreted as a constant $a$ and $v$ is interpreted as constant $b$, then we expect the evaluation engine to identify $a$ and $b$ in all contexts henceforth.
For example, the premise of the transitivity axiom $\mathrm{Le}(u, v) \land \mathrm{Le}(v, w) \implies \mathrm{Le}(u, w)$ should match tuples $(a, b_1), (b_2, c) \in \mathrm{Le}$ if an equality $b_1 = b_2$ has been inferred previously.

Partial functions can be encoded in RHL using relations representing the graphs of partial functions.
Thus one identifies partial functions $f : s_1 \times \dots \times s_n \rightarrow s$ with relations $f : s_1 \times \dots s_n \times s$ where the first $n$ components of each entry represent an element in the domain of the function and the last component represents the value of the function.
The \emph{functionality axiom}
\begin{equation}
  \label{eq:functionality}
  f(v_1, \dots, v_n, u) \land f(v_1, \dots, v_n, w) \implies u \equiv w
\end{equation}
enforces that the relation $f$ does indeed correspond to a well-defined partial function.

Sort quantifications $v : s$ in premises allow matching elements of a given sort $s$ that do not appear in any relation.
In standard Datalog, all variables in the head of a sequent must also appear in the body.
This requirement is removed in RHL.
Variables that only appear in the conclusion are implicitly existentially quantified.
If the premise of a sequent matches, then the evaluation engine must extend the match to the conclusion by finding an interpretation of variables that occur only in the conclusion such that the conclusion holds.
If no such extension exists, then the evaluation engine must create new identifiers to interpret variables that only occur in the conclusion, and enforce that the atoms of the conclusion hold for this interpretation.
We expect the evaluation engine to output a list of identifiers of each sort, including those identifiers that were created during evaluation.
An RHL sequent in which all conclusion variables also occur in the premise is called \emph{surjective}.
If an RHL theory contains non-surjective sequents, then evaluation need not terminate.
The Souffle Datalog engine implements a similar mechanism in the form of its choice construction \citep{souffle-choice}.

The presence of non-surjective sequents can not only lead to non-termination but also to non-deterministic results, in the sense that the result depends on the order in which sequents are matched.
This is not the case for \emph{strong} RHL theories, in which the interpretion of conclusion variables is uniquely determined once all sequents are satisfied.
For example, the RHL theory given by the functionality sequent \eqref{eq:functionality} and the sequent
\begin{equation}
  \label{eq:totality}
  v_1 : s_1 \land \dots \land v_n : s_n \implies f(v_1, \dots, v_n, v)
\end{equation}
is strong, since if the functionality axiom is satisfied, then the interpretation of the variable $v$ in sequent \eqref{eq:totality} is uniquely determined.
Unfortunately, it is undecidable whether a given RHL theory is strong.

A further problem with RHL is that functions must be encoded via their graphs.
This leads to excessive verbosity when formulating axioms involving complex expressions built up from function symbols.

\emph{Partial Horn logic} (PHL) \citep{phl} is a syntactic layer on top of RHL that rectifies these shortcomings.
In partial Horn logic, relations must be explicitly declared as predicates or partial functions.
Predicates correspond directly to RHL relations, whereas functions are lowered into a relation corresponding to the graph of the function and the implicit functionality axiom \eqref{eq:functionality}.

In positions where RHL accepts variables, PHL also allows composite terms
\begin{equation}
  t \, \text{::=} \, v \mid f(t_1, \dots, t_n)
\end{equation}
which are recursively defined from variables and application of partial function symbols to terms whose sorts match the function signature.
When lowering PHL to RHL, composite terms $t$ are recursively lowered into a result variable representing the term $t$ and additional RHL atoms.
These additional atoms are inserted into premise or conclusion of the sequent, depending on where $t$ appears.
To lower a composite term $t = f(t_1, \dots, t_n)$, we may assume that $t = f(v_1, \dots, v_n)$ for variables $v_i$ by recursively lowering the arguments $t_1, \dots, t_n$ first.
We now choose a fresh variable $v$ representing $t$ and add the RHL atom $f(v_1, \dots, v_n, v)$.
Since lowering a PHL formula reduces nested expressions into a flat sequence of atoms, the process of lowering PHL to RHL is also called \emph{flattening}.

PHL sequents are \emph{epic} if all variables in the conclusion are already introduced in the premise.
Note that lowering epic PHL sequents can result in non-surjective RHL sequents, because lowering composite terms can introduce fresh variables.
Nevertheless, the lowered RHL theory resulting from an epic PHL theory (i.e. a PHL theory containing epic sequents only) is strong.
Conversely, every strong RHL theory is equivalent to an epic PHL theory; see \citet[Section 4.3]{phl-theory} for details.
Thus, epic PHL has the same descriptive strength as strong RHL.
On the other hand, checking whether a PHL sequent is epic is trivial, whereas checking whether an RHL theory is strong is undecidable.
This makes PHL more suitable as human-facing input language to an evaluation engine compared to RHL.

The \emph{Eqlog} engine, whose underlying algorithm is described in this paper, accepts epic PHL as input language.
Eqlog lowers a user-provided epic PHL theory to RHL, which is then transpiled to a Rust module.
This is similar to the Souffle Datalog engine, which transpiles Datalog to C++.
In contrast to Souffle, Eqlog is meant to be used as part of a larger project and not as standalone tool.
Similarly to the \emph{egg} equality saturation library \citep{egg}, Eqlog supports online use-cases in which one alternates inserting new ground facts into the model and closing the model under PHL sequents.
Refer to the Eqlog project homepage \citep{eqlog-homepage} for details.

Independently of the work on Eqlog presented in this article, members of the Egg community have created a very similar tool that combines e-graphs with Datalog, which will be reported on in an upcoming article.

\textbf{Outline.}
In Section \ref{sec:rhl-evaluation}, we describe a basic algorithm to evaluate RHL, and a technique to detect termination in many circumstances.
Then, in Section \ref{sec:optimizations}, we discuss optimizations of the RHL evaluation algorithm.
Finally, in Section \ref{sec:applications}, we sketch applications of PHL and RHL evaluation to the implementation of programming languages.
Section \ref{sec:conclusion} concludes.

\textbf{Acknowledgements.}
Jakob Botsch Nielsen contributed significantly to previous versions of the implementation of the Eqlog tool and its application to an experimental type checker for a dependently typed proof assistant.
My own work on Eqlog and the algorithm presented in this paper commenced during my time as PhD student at Aarhus University, where I was advised by Bas Spitters and supported by the Air Force Office and Scientific Research project “Homotopy Type Theory and Probabilistic Computation”, grant number 12595060.

\section{RHL Evaluation}
\label{sec:rhl-evaluation}

In this section, we describe a basic algorithm to evaluate RHL theories.
The input to the evaluation algorithm is a list of RHL sequents and a \emph{relational structure} representing ground facts.
A relational structure is given by sets of numerical identifiers for each sort, representing the elements of the sort, and sets of tuples for each relation.
From Section \ref{subsec:naive-rhl-evaluation} onward, we assume that relational structures contain also union-find data structures for each sort, representing semantic equality of sort elements.

If RHL evaluation terminates, then we require that the output is a relational structure that is \emph{weakly free} with respect to the list of RHL sequents over the input relational structure \citep{phl-theory}.
Intuitively, this means that the output must satisfy all sequents in the RHL theory, and that it must be obtained from the input relational structure only from matching sequent premises and adjoining data corresponding to conclusions.

Weak freeness does not uniquely characterize a relational structure.
In general, the output relational structure depends on the order of matching premises of RHL sequents.
However, if the RHL theory is strong, then the output relational structure is \emph{(strongly) free} over the input relational structure, which determines it uniquely up to unique isomorphism (i.e. renaming of identifiers).
Relevant classes of strong RHL theories are theories containing surjective sequents only, and theories that are obtained from lowering epic PHL theories \citep{phl-theory}.

In Section \ref{subsec:naive-datalog}, we discuss the naive algorithm for Datalog evaluation and amend it with support for non-surjective sequents and sort quantification, but not equality.
Then, in Section \ref{subsec:naive-rhl-evaluation}, we consider a well-known simple congruence closure algorithm, which we shall understand as a special-purpose algorithm for evaluation of functionality RHL sequents.
The conclusion of functionality sequents is an equality, which our naive Datalog evaluation algorithm cannot process.
\emph{Union-find data structures} and \emph{normalization} are aspects of this congruence closure algorithm that deal with equalities in particular. 
We incorporate these into our naive Datalog algorithm to obtain a naive RHL evaluation algorithm.

In Section \ref{subsec:termination}, we discuss an example where our RHL evaluation algorithm does not terminate for a non-surjective RHL theory with finite free models.
Based on the example, we show how the evaluation algorithm can be modified to terminate for this particular example and also a wide range of other RHL theories.

\subsection{Naive Datalog evaluation}
\label{subsec:naive-datalog}

The naive Datalog algorithm is given by repeating \emph{premise matching} and \emph{conclusion application} phases until a fixed point is reached.
The high-level structure of the algorithm can be expressed in Rust-like pseudo-code as follows:
\begin{lstlisting}[language=rust,style=colouredRust]
fn datalog(structure, sequents) {
  loop {
    // 1. Match premises.
    let matches = [];
    for sequent in sequents {
      matches.push(find_matches(structure, sequent.premise));
    }

    // 2. Apply conclusions.
    let has_changed = false;
    for (sequent, matches) in sequents.zip(matches) {
      for match in matches {
        if apply_conclusion(structure, sequent.conclusion, match) {
          has_changed = true;
        }
      }
    }

    // Terminate if applying conclusions had no effect.
    if !changed {
      break;
    }
  }

  return structure;
}
\end{lstlisting}

\texttt{find\_matches} is a subprocedure that returns a list of matches of the given formula in a relational structure.
Each match is given by a mapping from the set of variables that occur in the formula to elements in the relational structure.
A naive implementation of this function enumerates matches using a nested loop join.
For example, matches of the formula $\mathrm{Le}(u, v) \land \mathrm{Ge}(w, v) \land \mathrm{Le}(w, x)$ can be enumerated as follows:
\begin{lstlisting}[language=rust,style=colouredRust]
  for (u, v) in structure.rels[Le] {
    for (w, v1) in structure.rels[Ge] {
      if v1 != v { continue; }
      for (w1, x) in structure.rels[Le] {
        if w1 != w { continue; }
        matches.push({u, v, w, x});
      }
    }
  }
\end{lstlisting}
Each relational atom translates into a nested loop over the corresponding relation in the relational structure, and each sort quantification translates into a loop over the corresponding list of elements.

\texttt{apply\_conclusion} is a subprocedure that inserts data into a relational structure according to a provided conclusion and a substitution of variables for elements in the relational structure.
It returns a boolean value indicating whether the operation had an effect, i.e. whether at least some of the concluded data was not already present in the relational structure.
For surjective sequents without equalities, where every variable in the conclusion is already bound by a match of the premise, we substitute the variables in each relation atom and insert the corresponding tuple into the relational structure.

For non-surjective sequents, we first check if the provided substitution of premise variables can be extended to interpretations of the conclusion variables such that the conclusion holds.
This can be accomplished using a version of the \texttt{find\_matches} function that takes a list of already fixed interpretations of some of the variables in the formula.
If no such extension exists, then we adjoin fresh elements to the relational structure to interpret the unbound conclusion variables and proceed as in the surjective case.

\subsection{Congruence closure and naive RHL evaluation}
\label{subsec:naive-rhl-evaluation}

RHL equality atoms can be reduced to Datalog by introducing binary equality relations on each sort representing inferred equality.
However, this \emph{setoid reduction} \citep[Section 3.4]{phl-theory} typically leads to inefficient Datalog programs.
Semantically, an inferred equality often reduces the size of the relational structure, since equalities can collapse two previously distinct tuples in a relation into a single tuple.
Instead, the setoid reduction leads to significant duplication due to congruence axioms, which assert that all relations must be closed in each argument under inferred equality.
Our goal in this section is to rectify this deficiency:
Every inferred equality should only shrink the relational structure.

Observe that RHL can be used to solve, in particular, the congruence closure problem:
Decide which equalities among a list $t_1, \dots, t_n$ of expression follow from a list of equalities among subexpressions.
This problem can be encoded in RHL with a theory given by an $(n + 1)$-ary relation symbol $f$ representing the graph of an $n$-ary function symbol that occurs in the $t_i$ and the \emph{functionality axiom}
\begin{equation}
  \label{eq:functionality}
  f(v_1, \dots, v_n, u) \land f(v_1, \dots, v_n, w) \implies u \equiv w.
\end{equation}
One then inserts data corresponding to the $t_i$ into a relational structure, imposes equalities among subexpressions, and closes the structure under functionality axioms.
We may thus understand congruence closure algorithms as special-purpose evaluation algorithms for functionality RHL sequents, and try to generalize existing congruence closure algorithms to general RHL evaluation.

Our inspiration here is the congruence closure algorithm described in \citet{congruence-closure}.
Consider the following version of their naive algorithm 2.1, which they attribute to \citet{naive-congruence-closure}.
The version presented here is specialized to a single binary function.
The input of the algorithm is a list of triples representing the graph of the function.
\begin{lstlisting}[language=rust,style=colouredRust]
  fn congruence_closure(graph) {
    uf = UnionFind::new();

    loop {
      // 1. Match premises.
      let eqs = [];
      for (x0, x1, x2) in graph {
        for (y0, y1, y2) in graph {
          if x0 == y0 && x1 == y1 {
            eqs.push((x2, y2));
          }
        }
      }

      // 2. Apply equalities.
      let has_changed = false;
      for (lhs, rhs) in eqs {
        lhs = uf.find(lhs);
        rhs = uf.find(rhs);
        if lhs != rhs {
          uf.union(lhs, rhs);
          has_changed = true;
        }
      }

      // Terminate if nothing has changed.
      if !has_changed {
        break;
      }

      // 3. Normalize.
      graph0 = [];
      for (x0, x1, x2) in graph {
        graph0.push(uf.find(x0), uf.find(x1), uf.find(x2));
      }
      graph = graph0;
    }

    return uf;
  }
\end{lstlisting}
Similarly to the Datalog evaluation algorithm of Section \ref{subsec:naive-datalog}, this congruence closure algorithm repeats a number of steps until it reaches a fixed point.
Step 1 corresponds to the \texttt{find\_matches} function for the premise of the functionality axiom \eqref{eq:functionality} of a binary function.

Step 2 applies the conclusions $u \equiv v$ for each match that was found in step 1.
The algorithm uses a union-find data structure to represent equality.
A union-find data structure associates a canonical representative to each equivalence class.
Equivalence classes are equal if and only if they have the same canonical representative.

Union-find data structures support fast \texttt{find} and \texttt{union} operations in near-constant runtime.
The \texttt{find} operation computes the canonical representative in the equivalence class of a given element.
The \texttt{union} operation merges the equivalence classes of two canonical representatives.

Step 3, which replaces all elements in entries of the \texttt{graph} relation by canonical representatives, does not have a counterpart in Datalog evaluation.
Because of the use of the union-find data structure, only comparisons among canonical representatives reflect inferred equality.
Note that, instead of the normalization step, we could also consult the union-find data structure in step 1 during premise matching when comparing elements.
However, a separate normalization step makes the use of a number of optimizations possible, which we discuss in Section \ref{sec:optimizations}.

By incorporating aspects of the congruence closure algorithm that deal with equalities into the naive Datalog evaluation algorithm of Section \ref{subsec:naive-datalog}, we now obtain our \emph{naive RHL evaluation algorithm}:
\begin{enumerate}
  \item
    In addition to sets of elements and relation, relational structures now contain also union-find data structures for each sort, representing semantic equality.
    We maintain the invariant that the relations in the relational structure contain canonical representatives only before each iteration of the evaluation algorithm.
  \item
    \texttt{apply\_conclusion} handles equalities $u \equiv v$ by merging the equivalence classes of the interpretations of $u$ and $v$ with a call to \texttt{union}.
  \item
    Before the end of the loop body, we insert a normalization step, which replaces each element in a tuple in any relation with its canonical representative by calling \texttt{find}.
\end{enumerate}
Since relational structures store relations as sets without duplication, normalization can potentially reduce the number of elements of relations.

The Souffle Datalog engine provides an efficient implementation of equivalence relations using union-find data structures \citep{souffle-union-find}, but it does not implement normalization.

\subsection{Detecting termination}
\label{subsec:termination}

If all sequents in the RHL theory to be evaluated are surjective, then the algorithm we have described in Section \ref{subsec:naive-rhl-evaluation} is guaranteed to terminate.
For non-surjective sequents, however, the (weakly) free model over a finite structure need not be finite.
In these situations, RHL evaluation can thus not terminate and must instead be aborted after a timeout is exceeded or a desired property has been inferred.
Nevertheless, there are RHL theories for which free models over finite relational structures are again finite despite the presence of non-surjective sequents.
But even in these situations, the RHL evaluation algorithm we have discussed so far need not terminate.

Consider, for example, the following PHL theory that axiomatizes pairs of maps $f : A \rightleftarrows B : g$ such that $g(f(x)) = x$ for all $x \in A$:
\begin{enumerate}
  \item
    \label{itm:f-total}
    $x : A \implies f(x) \downarrow$
  \item
    \label{itm:g-total}
    $y : B \implies g(y) \downarrow$
  \item
    \label{itm:retract}
    $y = f(x) \implies g(y) = x$
\end{enumerate}

Axiom \ref{itm:retract} is lowered to the RHL axiom $f(x, y) \implies g(y, x)$, which is surjective.
Axioms \ref{itm:f-total} and \ref{itm:g-total}, however, are non-surjective.
Nevertheless, free models of this theory over finite relational structures are always finite, as the following construction of free models proves:
Given sets $A, B$ and relations $f \subseteq A \times B, g \subseteq B \times A$, one first identifies elements within $A$ and $B$ according to the functionality axioms for $f$ and $g$ and axiom \ref{itm:retract}.
For each element $b \in B$ on which $g$ is not defined, we then adjoin new elements to $A$ and extend $g$ accordingly to a total function by axiom \ref{itm:g-total}.
Similarly, we then adjoin for each $a \in A$ on which $f$ is not defined new elements to $B$ and extend $f$ accordingly to a total function by axiom \ref{itm:f-total}.
Now every element in $B$ on which $g$ is not defined is of the form $g(a)$ for some unique $a \in A$, so by axiom \ref{itm:retract}, we may extend $g$ to a total function by setting $g(f(a)) = a$.
Now $f$ and $g$ are total functions and $g \circ f$ is the identity function on $A$.

On first thought, we might thus hope RHL evaluation for this theory to eventually reach a fixed point and terminate.
Unfortunately, this is not the case for the RHL evaluation algorithm described in Section \ref{subsec:naive-rhl-evaluation}.
Consider the iterations of evaluation with initial relational structure given by $A = \{ a_0 \}$, $B = \emptyset$ and $f$ and $g$ entirely undefined:
\begin{enumerate}
  \item
    Axiom \ref{itm:f-total} matches on $a_0 \in A$, resulting in a new element $b_0 \in B$ and the tuple $(a_0, b_0) \in f$.
  \item
    Axiom \ref{itm:g-total} matches on $b_0 \in B$, and axiom \ref{itm:retract} matches on $(a_0, b_0)$.
    The former results in a new element $a_1 \in A$ and the tuple $(b_0, a_1) \in g$, while the latter results in the tuple $(b_0, a_0) \in g$.
  \item
    \label{itm:infinite-f-adjoining}
    Axiom \ref{itm:f-total} matches on $a_1 \in A$, and the implicit functionality axiom for $g$ matches on $(b_0, a_0), (b_0, a_1)$.
    The former results in a new element $b_1 \in B$ and the tuple $(a_1, b_1) \in f$, while the latter results in the equality $a_0 = a_1$.
  \item
    \label{itm:infinite-g-adjoining}
    Axiom \ref{itm:g-total} matches on $b_1 \in B$, and the implicit functionality axiom for $f$ matches on $(a_0, b_0), (a_1, b_1)$.
    The former results in a new element $a_2 \in A$ and the tuple $(b_1, a_2) \in g$, while the latter results in the equality $b_0 = b_1$.
  \item
    All further iterations alternate between variations of iterations \ref{itm:infinite-f-adjoining} and \ref{itm:infinite-g-adjoining}.
\end{enumerate}
What prevents termination for this theory is thus that the evaluation algorithm matches all sequents at once:
Observe that our proof that free models are finite relies on applying sequents in some particular order.
For this particular theory, non-termination can be prevented by carefully stating axioms so as to avoid alternating states, for example by replacing axioms \ref{itm:f-total} and \ref{itm:retract} with $x : A \implies g(f(x)) = x$.

However, the following variant of the RHL evaluation algorithm described in Section \ref{subsec:naive-rhl-evaluation} terminates on a wide range of RHL theories, including the theory above.
One splits the top-level evaluation loop into an inner loop responsible for surjective sequents and an outer loop responsible for non-surjective sequents.
The algorithm thus alternates closing the relational structure under all surjective sequents (which always terminates) and a single step of matching and adjoining conclusions of non-surjective sequents.
If eventually a non-surjective step does not change the relational structure, then all sequents are satisfied and evaluation terminates.

However, I expect there to be theories where termination depends on a particular order in which non-surjective sequents are applied, and then this simple approach does not suffice.

\section{Optimizations}
\label{sec:optimizations}

In this section, we consider optimizations of the naive RHL algorithm that we discussed in Section \ref{subsec:naive-rhl-evaluation}.
Most of these techniques are adapted from optimizations that apply to Datalog evaluation, to the congruence closure problem or to both.
Implemented together, these optimizations allow us to recover the fast congruence closure algorithm due to \citet{congruence-closure} as a special case of RHL evaluation for functionality axioms.

\subsection{Semi-naive evaluation}
\label{subsec:semi-naive}

Semi-naive evaluation is a common Datalog evaluation optimization.
It exploits the observation that matches of premises that were found in the previous iteration need not be considered again because conclusions to these matches have already been adjoined.
A match has not been found in a previous iteration if at least one of the atoms in the premise is matched with new data, i.e. data was added only in the last iteration.
To distinguish old data from new data, we store for each relation and each sort lists of tuples or elements that were added in the last iteration. 
An $n$-fold nested loop that matches the premise of a sequent can now be replaced with $n$ copies, where in the $i$th copy the $i$th loop iterates over new data only.
For example, the nested loop described in Section \ref{subsec:naive-datalog} enumerating the premise of $\mathrm{Le}(u, v) \land \mathrm{Ge}(w, v) \land \mathrm{Le}(w, x)$ can be replaced by the following three nested loops:
\begin{lstlisting}[language=rust,style=colouredRust]
  for (u, v) in structure.rels_new[Le] {
    for (w, v1) in structure.rels_all[Ge] {
      if v1 != v { continue; }
      for (w1, x) in structure.rels_all[Le] {
        if w1 != w { continue; }
        matches.push([u, v, w, x]);
      }
    }
  }
  for (u, v) in structure.rels_all[Le] {
    for (w, v1) in structure.rels_new[Ge] {
      if v1 != v { continue; }
      for (w1, x) in structure.rels_all[Le] {
        if w1 != w { continue; }
        matches.push([u, v, w, x]);
      }
    }
  }
  for (u, v) in structure.rels_all[Le] {
    for (w, v1) in structure.rels_all[Ge] {
      if v1 != v { continue; }
      for (w1, x) in structure.rels_new[Le] {
        if w1 != w { continue; }
        matches.push([u, v, w, x]);
      }
    }
  }
\end{lstlisting}
Observe that not only the conclusion application phase but also the normalization phase can lead to new data:
If an element in the tuple of some relation changes as a result of normalization, then the tuple must be considered as new tuple.

The optimized congruence closure algorithm described by \citet{congruence-closure} also implements semi-naive evaluation.
Their \texttt{pending} list in algorithm 2.4 corresponds to our set of new tuples in the relation representing the graph of a function.

Semi-naive evaluation is well-suited for online applications, where one alternates RHL evaluation and ad-hoc manipulation.
If this manipulation consists only of adjoining data, then this data can be adjoined to the same data structures that hold new data during RHL evaluation.
The first iteration of subsequent RHL evaluation need then only consider matches for this new data instead of matches in the full relational structure.

\subsection{Symmetries}
\label{subsec:symmetries}

Semi-naive matching of the premise of the functionality axiom for a (binary) function results in two loops:
\begin{lstlisting}[language=rust,style=colouredRust]
  for (x0, x1, x2) in structure.rels_new[f] {
    for (y0, y1, y2) in structure.rels_all[f] {
      ...
    }
  }
  for (x0, x1, x2) in structure.rels_all[f] {
    for (y0, y1, y2) in structure.rels_new[f] {
      ...
    }
  }
\end{lstlisting}
On the other hand, the congruence closure algorithm described by \citet{congruence-closure} requires a single loop only.
Indeed, the second loop is unnecessary due to a \emph{symmetry} in the functionality axiom $f(v_0, v_1, u) \land f(v_0, v_1, w) \implies u \equiv w$.
The symmetry is given by swapping $u$ and $w$.
This results in a semantically equivalent premise, and swapping the variable has the same effect as swapping the two premise atoms.
In such cases, it suffices to consider matches where the first of the two atoms is interpreted with new data.
Another example where symmetries can be exploited is the anti-symmetry axiom $\mathrm{Le}(u, v) \land \mathrm{Le}(v, u) \implies u \equiv v$.

\subsection{Indices and occurrence lists}
\label{subsec:indices}

Indices are meant to speed up the nested loops that enumerate matches of premises.
The idea is to replace each inner loop by an efficient sublinear lookup with fixed projections.
For example, matching the premise $\mathrm{Le}(u, v) \land \mathrm{Le}(v, w)$ of a transitivity axiom can be sped up with an index on the first projection.
One thus maintains a map that allows fast lookup of all tuples $(v, w)$ for fixed $v$.
The premise can then be enumerated as follows:
\begin{lstlisting}[language=rust,style=colouredRust]
  for (u, v) in structure.rels[Le] {
    for (_, w) in structure.rels[Le].index[v] {
      matches.push((u, v, w));
    }
  }
\end{lstlisting}
Indices are typically realized using variants of ordered search trees or hash maps.
They can be maintained over all iterations, or recreated before and disposed immediately after the premise matching phase in each iteration.
Recreating indices requires less memory compared to maintaining indices, since only indices needed to match a single sequent need to be stored in memory at any time.

Fast Datalog engines often maintain indices over all iterations, which results in faster execution at the expense of increased memory usage.
If indices are maintained over all iterations, new tuples must also be inserted into indices during the conclusion application phase.
For RHL evaluation, however, index maintenance is problematic due to the normalization phase, in which elements in relations are replaced by their canonical representatives if needed.
When using indices, they require normalization, too.

We turn again to the fast congruence closure algorithm described by \citet[Section 2.4]{congruence-closure} to implement index normalization efficiently.
Their \emph{signature table} is a hash map index that speeds up matching premises of functionality axioms.
It is maintained throughout the iterations of the congruence closure algorithm.
Efficient normalization is implemented using further data structures which we shall refer to as \emph{occurrence lists}.
An occurrence list contains for each equivalence class the expression nodes which have at least one child in the equivalence class.
In the normalization step, it suffices to normalize those tuples that occur in occurrence lists of elements that ceased to be canonical representatives.

Occurrence lists can be adapted to RHL evaluation as follows.
We associate to each canonical representative a list of tuples in any relation in which the canonical representative occurs.
Tuples in occurrence lists need not be normalized.

In the conclusion application phase, we insert tuples also in the occurrence lists of each element of the inserted tuple.
When merging two equivalence classes, we save the element that ceases to be a canonical representative along with its occurrence list for use during the following normalization phase.
The occurrence lists of the element that remains a canonical representative is set to the concatenation of the two original occurrence lists.
Concatenation can be implemented asymptotically efficiently if occurrence lists are realized as linked lists or rope data structures.
In the normalization phase, we remove each tuple in one of the occurrence lists we saved earlier, normalize the tuple and reinsert it into each index.

When enforcing an equality during the conclusion phase, the algorithm described  in \citet[Section 2.4]{congruence-closure} chooses the element that remains a canonical representative in a way that minimizes the amount of normalization necessary:
Thus, the element with longer occurrence lists should remain canonical.
This applies directly also to occurrence lists in RHL evaluation.

To avoid normalizing tuples that were inserted in the current iteration, the conclusion application phase can be split into an \emph{equality application phase}, where we only consider equalities in conclusions, and a \emph{relation application phase}, where we only consider relation atoms.
We then normalize between the equality application phase and the relation application phase.
This has the benefit that new tuples need not be normalized directly after insertion.

\subsection{Functional projections}

We say that the $i$th projection of a relation $r : s_1 \times \dots \times s_n$ in an RHL theory is \emph{functional} if the $i$th projection $x_i$ of each tuple $(x_1, \dots, x_n) \in r$ is uniquely determined by the other components $x_1, \dots, x_{i - 1}, x_{i + 1}, \dots, x_n$ of the tuple.
More generally, we can consider a set $I$ of projections which are uniquely determined by the complementary projections.
As the name suggests, the functionality axiom for an $n$-ary function asserts that the $(n + 1)$th projection of the graph of the function is functional.
Another example are injective functions, where the first $n$ projections of the graph depend functionally on the $(n + 1)$th projection.

When indices are maintained on a relation with a functional projection, equality constraints can be generated already during insertion into the index instead of later during the matching phase.
For example, if $r$ is an $(n + 1)$-ary relation representing the graph of a function, then an index on the first $n$ arguments can be maintained to match the premise of the functionality axiom.
Without consideration of functionality of the $(n + 1)$th projection, we expect the index to allow lookups of \emph{lists} of tuples for fixed value of the first $n$ projections.
Due to the functional dependency, we can instead enforce that lookups result into at most one tuple.
Whenever a second tuple would be inserted into the index which would violate this property, then we generate equality constraints according to functional projections instead.
These equalities are then enforced during the next conclusion application phase.

The \emph{signature table} hash map in the efficient congruence closure algorithm described by \citet[Section 2.4]{congruence-closure} can be understood as an index with functional projection optimization.

\section{Applications}
\label{sec:applications}

In this Section, we discuss two applications of RHL and PHL evaluation to the implementation of programming languages:
Steensgaard's points-to analysis (Section \ref{subsec:steensgard}) and type inference (Section \ref{subsec:inference}).

\subsection{Steensgaard's points-to analysis}
\label{subsec:steensgard}

Points-to-analysis aims to bound the set of heap objects a variable in a program can point to.
The two well-known approaches are due to Andersen and Steensgaard.
Both algorithms identify heap objects with their allocation sites, i.e. the program expressions that allocate objects.
The Andersen and Steensgaard analyses thus give an over-approximating answer to the question of whether a given variable $x$ can point to an object that was allocated in expression $e$.

Both analyses must consider the case where a variable $x$ is a assigned to a variable $y$.
Andersen-style analysis computes for each variable $x$ a set of objects $e$ that $x$ can point to such that the following properties hold:
\begin{enumerate}
  \item
    If there is a statement that assigns an allocating expression $e$ to a variable $x$, then $x$ can point to $e$.
  \item
    If a variable $y$ can take the value of a variable $x$ (e.g. as a result of an assignment or a function call), and $x$ can point to $e$, then $y$ can point to $e$.
\end{enumerate}
A minimal implementation of Andersen-style analysis using Datalog populates input relations $\mathrm{Alloc}(x, e)$ and $\mathrm{Assign}(x, y)$ with data derived from the program source code.
The rules above governing the $\mathrm{PointsTo}(x, e)$ relation are then encoded in Datalog as follows:
\begin{enumerate}
  \item
    $\mathrm{Alloc}(x, e) \implies \mathrm{PointsTo}(x, e)$
  \item
    \label{itm:andersen-subset}
    $\mathrm{Assign}(x, y) \land \mathrm{PointsTo}(x, e) \implies \mathrm{PointsTo}(y, e)$
\end{enumerate}
To summarize, Andersen's algorithm enforces \emph{subset constraints}:
If a variable $y$ can take the value of a variable $x$, either through a function call or a direct assignment, then the points-to set of $x$ is a subset of the points-to set of $y$.

Steensgaard's algorithm is a less precise but typically faster variation of Andersen's algorithm.
It is based on \emph{equality constraints}.
Thus if a variable $y$ can take the value of a variable $x$, then Andersen's algorithm \emph{equates} the points-to sets of $x$ and $y$.
A direct implementation of Steensgaard's algorithm maintains a union-find data structure on the variables of a program, and a mapping that assigns to each canonical representative a points-to set of heap allocations.
The algorithm scans the program source code and performs the following actions:
\begin{enumerate}
  \item
    For each statement that assigns an allocation expression $e$ to a variable $x$, add $e$ to the points-to set of the canonical representative of $x$.
  \item
    If a variable $y$ can take the value of a variable $x$, unify the equivalence classes of $x$ and $y$.
    The points-to set of the unified equivalence class is the union of the points-to sets of the classes of $x$ and $y$.
\end{enumerate}
Steensgaard's algorithm is strictly less precise than Andersen's, but it typically requires less memory, since only one points-to set needs to be maintained for every equivalence class of variables.
To encode Steensgaard's algorithm in RHL, we can simply replace rule \ref{itm:andersen-subset} above in Andersen's analysis with the rule
\begin{equation}
  \mathrm{Assign}(x, y) \implies x \equiv y
\end{equation}
to enforce equality constraints.

\subsection{Type inference}
\label{subsec:inference}

Type inference (or type reconstruction) is the task of assigning types to variables and expressions based on their usage in a program fragment.
The constraint-based typing algorithm for the simply typed lambda calculus described in \citet[22.3, 22.4]{pierce-types-and-programming-languages} assigns to each term a separate, initially unrestricted type variable.
It then collects constraints on type variables according to the usage and definition of the corresponding terms.
This is accomplished by considering typing rules and their inverses.
For example, based on the application typing rule
\begin{equation}
  \label{eq:application-typing-rule}
  \inferrule
  { k : \sigma \rightarrow \tau \and s : \sigma }
  { k \, s : \tau }
\end{equation}
we infer the following constraints from a term $t_2 = t_0 \, t_1$:
\begin{enumerate}
  \item
    \label{itm:app-constraints-first}
    If $t_0$ has type $S \rightarrow T$, then $t_1$ has type $S$.
  \item
    \label{itm:infer-dom-from-arg}
    Conversely, if $t_1$ has type $S$, then $t_0$ has type $S \rightarrow T$ for some $T$.
  \item
    If $t_0$ has type $S \rightarrow T$, then $t_2$ has type $T$.
  \item
    \label{itm:infer-cod-from-result}
    Conversely, if $t_2$ has type $T$, then $t_0$ has type $S \rightarrow T$ for some $S$.
\end{enumerate}
The implicit existentials in constraints \ref{itm:infer-dom-from-arg} and \ref{itm:infer-cod-from-result} generate new type variables $T$ and $S$, which must be chosen fresh for each instance of the constraint.
In the following unification step, the generated constraints are checked for compatibility, and if so the most general substitution of type variables is created that satisfies all the constraints.
For example, for the fragment $x \, y$ for variables $x$ and $y$, the algorithm outputs the substitution $[T_0 \mapsto (T_1 \rightarrow T_2), T_1 \mapsto T_1, T_2 \mapsto T_2]$, where $T_0, T_1$ and $T_2$ are the type variables initially assigned to $x, y$ and $x \, y$.

This inference procedure can be implemented in PHL as follows.
We introduce sorts $\mathrm{Tm}$ of terms and $\mathrm{Ty}$ of types.
Each $n$-ary type or term constructor corresponds to an $n$-ary PHL function.
For example, function types are encoded as binary function symbol $\mathrm{Fun} : \mathrm{Ty} \times \mathrm{Ty} \rightarrow \mathrm{Ty}$ and application as a function $\mathrm{Tm} \times \mathrm{Tm} \rightarrow \mathrm{Tm}$.
To enforce injectivity of type constructors, we introduce inverse functions for each parameter of a type constructor.
For function types, we add functions $\mathrm{Dom}: \mathrm{Ty} \rightarrow \mathrm{Ty}$ and $\mathrm{Cod} : \mathrm{Ty} \rightarrow \mathrm{Ty}$ and enforce axioms that $\mathrm{Dom}$ and $\mathrm{Cod}$ are indeed inverses to $\mathrm{Ty}$:
\begin{mathpar}
  \mathrm{Dom}(\kappa) \downarrow \implies \mathrm{Cod}(\kappa) \downarrow \and
  \mathrm{Cod}(\kappa) \downarrow \implies \mathrm{Dom}(\kappa) \downarrow \\
  \sigma = \mathrm{Dom}(\kappa) \land \tau = \mathrm{Dom}(\kappa) \implies \kappa = \mathrm{Fun}(\sigma, \tau) \\
  \kappa = \mathrm{Fun}(\sigma, \tau) \implies \mathrm{Dom}(\kappa) = \sigma \land \mathrm{Cod}(\kappa) = \tau
\end{mathpar}
Thus $\mathrm{Dom}$ and $\mathrm{Cod}$ are defined on the same set of types: Those of the form $\mathrm{Fun}(\sigma, \tau)$ for types $\sigma$ and  $\tau$.

To detect violations of joined injectivity of type constructors, we introduce a nullary predicate $\bot$ and rules such as
\begin{mathpar}
  \mathrm{Fun}(-, -) \equiv \mathrm{List}(-) \implies \bot()
\end{mathpar}
for every pair of distinct type constructors, for example function types and list types.
We always arrange $\bot$ to be empty before PHL evaluation, so that $\bot$ is inhabited after evaluation if and only if a violation of joined injectivity was inferred.

We encode the typing relation $t : \tau$ as a function $\mathrm{TmTy} : \mathrm{Tm} \rightarrow \mathrm{Ty}$ instead of a relation because each term has a unique type.
The axiom $t : \mathrm{Tm} \implies \mathrm{TmTy}(t) \downarrow$ enforces that each term has a type.
During evaluation, this non-surjective rule introduces a fresh identifier as type of each term if necessary.

Finally, we encode term constructors as PHL functions and add axioms according to inference rules and their inverses.
For example, the typing rule \eqref{eq:application-typing-rule} of function application results in a PHL function $\mathrm{App} : \mathrm{Tm} \times \mathrm{Tm} \rightarrow \mathrm{Tm}$ and the following PHL axioms governing it, corresponding to the constraints \ref{itm:app-constraints-first} -- \ref{itm:infer-cod-from-result} above:
\begin{enumerate}
  \item
    $\mathrm{App}(t_0, t_1) \downarrow \land \mathrm{TmTy}(t_0) = \mathrm{Fun}(\sigma, -) \implies \mathrm{TmTy}(t_1) = \sigma$
  \item
    \label{itm:axiom-dom-exists}
    $\mathrm{App}(t_0, t_1) \downarrow \land \mathrm{TmTy}(t_1) = \sigma \implies \mathrm{Dom}(\mathrm{TmTy}(t_0)) = \sigma$
  \item
    $t_2 = \mathrm{App}(t_0, t_1) \land \mathrm{TmTy}(t_0) = \mathrm{Fun}(-, \tau) \implies \mathrm{TmTy}(t_2) = \tau$
  \item
    \label{itm:axiom-cod-exists}
    $t_2 = \mathrm{App}(t_0, t_1) \land \mathrm{TmTy}(t_2) = \tau \implies \mathrm{Cod}(\mathrm{TmTy}(t_0)) = \tau$
\end{enumerate}
Observe that the conclusions of axioms \ref{itm:axiom-dom-exists} and \ref{itm:axiom-cod-exists} assert that domains and codomains of certain types exist, which together with our axioms for $\mathrm{Dom}$ and $\mathrm{Cod}$ implies that these types are function types.
If necessary, fresh type identifiers are adjoined during PHL evaluation by the non-surjective axioms asserting that the $\mathrm{Dom}$ and $\mathrm{Cod}$ functions are defined on the same set of types.

With the PHL theory modeling the type system at hand, the full type inference algorithm can now be implemented in three steps:
\begin{enumerate}
  \item
    Populate a relational structure based on the program fragment:
    We adjoin a unique term identifier for each term in the program fragment and add entries in the relations representing graphs of term constructors.
  \item
    Close the relational structure under the axioms described above.
  \item
    If the $\bot$ relation contains an element, output an error.
    Otherwise, there exists for each type identifier $T$ at most one type constructor $\mu$ and an entry of the form $(T_1, \dots, T_n, T)$ with $T$ as last component in the relation corresponding to $\mu$.
    Expand each type identifier recursively into a maximal syntax tree according to such entries.
    Output the maximal syntax tree representing $\mathrm{TmTy}(t)$ for each term $t$ in the input program fragment.
\end{enumerate}

\section{Conclusion}
\label{sec:conclusion}

Using Datalog to implement type checking and inference, as we sketched in Section \ref{subsec:inference}, is not a new idea.
For example, the blog post \citet{lowering-rust-traits-to-logic} discusses an implementation of Rust's trait system using Datalog.
One of the issues raised there is that Rust's associated types require reasoning about equalities of types.
A similar issue would arise for Haskell's type families.
The solution proposed in the blog post is to combine Datalog with a normalization algorithm.
RHL's built-in equality might offer a declarative alternative to normalization that applies also in situations where no strongly normalizing rewrite system is available.

Typing rules are typically specified using the notation of natural deduction (e.g. the application typing rule \eqref{eq:application-typing-rule}).
Apart from syntactic differences, the structure of such rules and their semantics correspond almost precisely to PHL.
Indeed, it is generally understood that many type systems can be encoded as \emph{essentially algebraic theories} \citep[Chapter 3.D]{locally-presentable-and-accessible-categories}, which have the same descriptive strength as epic PHL.
From this perspective, program fragments can be identified with elements of the \emph{initial model} of the PHL theory axiomatizing the type system, i.e. the free model over the empty relational structure.

These conceptual connections and the example in Section \ref{subsec:inference} suggest that PHL and RHL evaluation have the potential to assume a role in the implementation of programming languages paralleling that of parser generators:
Where parser generators produce parsers from grammars, RHL evaluation engines produce type checkers from type systems.

\bibliographystyle{abbrvnat}
\setcitestyle{authoryear,open={(},close={)}}
\bibliography{main}

\end{document}